\documentclass{ws-procs9x6}

\usepackage{graphicx}

\newcommand{\beq}{\begin{eqnarray}}
 \newcommand{\eeq}{\end{eqnarray}}
\newcommand{\be}{\begin{equation}}
 \newcommand{\ee}{\end{equation}}

 \def\la{\mathrel{\mathpalette\fun <}}

\def\fun#1#2{\lower3.6pt\vbox{\baselineskip0pt\lineskip.9pt
\ialign{$\mathsurround=0pt#1\hfil ##\hfil$\crcr#2\crcr\sim\crcr}}}

\newcommand{{\SD}}{\rm SD}
\newcommand{{\Lc}}{\mathcal{L}}
\newcommand{{\Mc}}{\mathcal{M}}

\newcommand{\ver}{\mbox{\boldmath${\rm r}$}}
\newcommand{\vesig}{\mbox{\boldmath${\rm \sigma}$}}
\newcommand{\venab}{\mbox{\boldmath${\rm \nabla}$}}

\newcommand{\veP}{\mbox{\boldmath${\rm P}$}}
\newcommand{\vep}{\mbox{\boldmath${\rm p}$}}

\newcommand{\veA}{\mbox{\boldmath${\rm A}$}}

\newcommand{\veK}{\mbox{\boldmath${\rm K}$}}
\newcommand{\veR}{\mbox{\boldmath${\rm R}$}}

\newcommand{\vexi}{\mbox{\boldmath${\rm \xi}$}}
\newcommand{\vepi}{\mbox{\boldmath${\rm \pi}$}}
\newcommand{\veta}{\mbox{\boldmath${\rm \eta}$}}
\newcommand{\veB}{\mbox{\boldmath${\rm B}$}}

\newcommand{\ran}{\rangle}

\begin{document}

\title{Composite systems in magnetic field: from hadrons to hydrogen atom}
\author{B. O.~Kerbikov$^*$}
\address{Institute of Theoretical and Experimental Physics, B.~Cheremushkinskaya, 25,\\
Moscow, 117218 Russia\\
$^*$E-mail: borisk@itep.ru\\
www.itep.ru}
\address{Moscow Institute of Physics and technology, Institutesky line 9,\\
Moscow region, Dolgoprudny, 141701 Rusia\\
www.mipt.ru}

\begin{abstract}
We briefly review the recent studies of the behavior of  composite systems in
magnetic field. The hydrogen atom is chosen to demonstrate the new results
which may be experimentally tested. Possible applications to physics of
antihydrogen are mentioned.
\end{abstract}


\bodymatter

\section{Introduction}

We are witnessing an outburst of interest to the behavior of quantum systems in
strong magnetic field (MF) \cite{1}. This is probably due to the fact that huge
MF up to $eB\sim \Lambda^2_{QCD} \sim 10^{19} G$ has become a physical reality.
Such field is created  (for  a short time) in heavy ion collisions at RHIC and
LHC \cite{2}. The field about four orders of magnitude less is anticipated to
operate in magnetars \cite{3}. It is impossible in a brief presentation to
cover the results of  intensive studies performed by an impressive number of
researches. I concentrate mainly on the work of ITEP group (M.~A.~Andreichikov
B.~O.~Kerbikov, V.~D.~Orlovsky, and Yu.~A.~Simonov). And even more concise on the
hydrogen atom in MF problem. The results concerning  the quark systems in MF
will be merely formulated.

The quantum mechanics of charged particle in magnetic field is presented in
textbooks \cite{4}. In a constant  MF assumed to be along the $z$ axis the
transverse motion is quantized into Landau levels $(\hbar =c=1)$

\be E_\bot = \left( n +\frac12\right) \omega_H,~~ n=0,1,2,...,\label{1}\ee
where $\omega_H= \frac{|e|B}{m}$ is the cyclotron frequency. The quantity which
in MF takes the role of the mechanical momentum, commutes with the Hamiltonian,
and is therefore a constant  of motion, is  a \underline{pseudomomentum}
\cite{4,5,6,7,8} \be \hat{\veK}= \vep - e \veA + e\veB \times \ver =  - i
 \venab - e \veA + e \veB \times \ver.\label{2}\ee

In the London gauge $\veA(\ver) = \frac12 \veB\times \ver$ the pseudomomentum
takes the form \be \hat{\veK} = \vep +\frac12 \veB\times \ver.\label{3}\ee

Mathematically, the conservation of $\veK$ reflects the invariance under the
combined action of the spatial translation and the gauge transformation.
Physically, $\veK$ is  conserved since it takes into account the Lorentz force
acting on a particle in MF (motional electric field).

The importance of pseudomomentum becomes clear when we turn to a two-body, or
many-body problems in MF.

\section{The wave function factorization in MF}

The total momentum of $N$ mutually  interacting particles with translation
invariant interaction is a constant of motion and the center of mass motion can be
separated in the Schrodinger equation. For the system with total electric
charge $Q=0$ embedded in MF factorization of the wave function can be performed
making use of the pseudomomentum operator $\hat K$ \cite{5,6,7,8}. As a simple
example consider two nonrelativistic particles with masses $m_1$ and $m_2$,
charges $e_1=e>0, e_2 =-e$, and interparticle interaction $V(\ver_1-\ver_2)$.
The hydrogen atom is such a system. The Hamiltonian reads \be \hat
H=\frac{1}{2m_1} (\vep_1 - e\veA(\ver_1))^2 + \frac{1}{2m_2} (\veP_2 + e\veA
(\ver_2))^2 + V(\ver_1-\ver_2).\label{4}\ee Choosing the gauge $\veA=\frac12
\veB \times \ver$ and introducing $$ M=m_1+m_2, \mu=m_1m_2 (m_1+m_2)^{-1}, s=
(m_1-m_2)(m_1+m_2)^{-1},$$

$$ \ver=\ver_1-\ver_2, \veR = (m_1\ver_1 + m_2\ver_2) M^{-1},
\veP=-i\partial/\partial\veR,$$ $\vepi =- i\partial/\partial\ver$,  we obtain
\be \hat H=\frac{1}{2M} (\veP +\frac{e}{2} \veB \times \ver)^2 + \frac{1}{2\mu}
(\vepi + \frac{e}{2} \veB \times \veR + s \frac{e}{2} \veB\times
\ver)^2.\label{5}\ee

The two-body pseudomomentum operator is \be \hat{\veK} = \sum^2_{i=1} (\vep_i +
\frac12 e_i \veB\times \ver_i) = - i \frac{\partial}{\partial \veR} -
\frac{e}{2} \veB \times \ver.\label{6}\ee

Since $\hat{ \veK}$ commutes with $\hat H,$ the full two-particle wave function
$\Psi (\veR, \ver)$ is the eigenfunction of $\hat {\veK} $ with the eigenvalue
$\veK$

\be \hat {\veK} \Psi (\veR, \ver) = \veK \Psi (\veR, \ver),\label{7}\ee

The wave  function which satisfies (\ref{7}) has the form \be \psi(\veR, \ver)
=\exp \{ (\veK + \frac{e}{2} \veB \times \ver) \veR\} \varphi
(\ver).\label{8}\ee

Substitution of the ansatze (\ref{8}) into $\hat H$ leads to the equation \be
\left[ \frac{\veK^2}{2 M} + \frac{e}{M} (\veK \times \veB) \ver+
\frac{\vepi^2}{2\mu} + \frac{e}{2} \frac{s}{\mu} \veB (\ver \times \vepi) +
\frac{e^2}{8\mu} (\veB \times \ver)^2 + V(\ver) \right] \varphi_K (\ver) = E
\varphi_K (\ver).\label{9}\ee The subscript $K$ affixed to $E$ reflects the fact that
$E$ has a residual dependence on $\veK$ through the second term in (\ref{9}).

For harmonic interaction $V(\ver) = \frac{\sigma^2}{m}\ver^2$ the problem has
an analytical solution and the ground state energy corresponds to $\veK=0$. The
simple calculation yields $(m_1=m_2=m)$ \be E=2\Omega (n_x+ n_y +1) + 2 \omega
(n_2 + \frac12) + \frac{1}{4m} \left[
\frac{K^2_x+K_y^2}{1+\left(\frac{eB}{2\sigma}\right)^2} +
K^2_z\right],\label{10}\ee \be \Omega= \frac{\sigma}{m}
\sqrt{1+\left(\frac{eB}{2\sigma}\right)^2}, \omega
=\frac{\sigma}{m}.\label{11}\ee

To complete this section, we present examples of the pseudomomentum for three-
and  four-body systems. Consider a model of the neutron as a system of two $d$-quarks with charges  -$e/3$ and masses $m_d$, and  one $u$-quark with a
charge $2e/3$ and a mass $m_u, M=2 m_d + m_u$. This problem was formulated in Ref.~\refcite{9} and is now under investigation in relativistic formalism. Following Ref.~\refcite{9} we introduce the Jacobi coordinates.

\be \veta = \frac{\ver_1 -\ver_2}{\sqrt{2}}, \vexi = \sqrt{\frac{m_u}{2M}}
(\ver_1 + \ver_2 - 2 \ver_3), \veR=\frac{1}{M} \sum^3_{i=1} m_i
\ver_i.\label{12}\ee Then \be \hat{ \veK} = \sum^3_{i=1} (\vep_i +\frac12 e_i
\veB\times\ver_i) = \veP + \frac{e}{2} \sqrt{\frac{M}{2m_u}} \veB \times
\vexi.\label{13}\ee As an example of a neutral four-body system consider
hydrogen-antihydrogen $H-\bar H$ \cite{10}. Let $\ver_1$ and $\ver_2$ be the
coordinates of $p$ and $\bar p$, $\ver_3$ and $\ver_4$  be the coordinates of $
e^-$ and $e^+$. Then \be \hat{ \veK} =  \veP + \frac{e}{2} \veB \times \{
(\ver_1-\ver_2) + (\ver_4 -\ver_3)\}=\veP + \frac{e}{2} \veB \times \{
(\ver_1-\ver_3) + (\ver_2 -\ver_4)\}.\label{14}\ee

This obvious result corresponds to the two possible configurations of the
system: a)$ p \bar p +  e^+e^-, ~~$  b)$ H-\bar H.$ Transitions between these
two configurations in MF as a Landau--Zener effect will be  a subject of a
forthcoming publication.

We have reminded the essential formalism needed to treat the composite system
under  MF. Now we turn to some physical problems.

\section{The Hierarchy of MF}

The present interest to the effects induced by MF  was triggered by the
realization of the fact that MF generated in heavy ion collisions reaches the
value $eB\sim 10^{19} G \sim \Lambda^2_{QCD}$. The highest MF which can be
generated now  in the laboratory is about $10^6 G$. From the physical point of
view there are two characteristic values of  MF strength. The Schwinger one is
$B_{cr} = m^2_e/e = 4.414\cdot 10^{13} G$. At $B=B_{cr}$ the distance between
the lowest Landau level (LLL) of the electron and the next one is equal to
$m_e$. This can be seen from (\ref{1}), or from the relativistic dispersion
relation  \be \omega_{n,\sigma} (p_z) = [p^2_z + m^2 + q B
(2n+1+\sigma)]^{1/2}.\label{15}\ee

Here MF is pointing along the $z$-axis, $m$ is the particle mass, $q$ is the
absolute value of its electric charge, $\sigma = \pm 1$ depending on the spin
projection. The LLL corresponds to $n=0$, $ \sigma =-1$. The second important
benchmark  is the atomic field $B_a = m_e^2 e^3 = 2.35\cdot 10^9 G$. At $B=B_a$
the Bohr radius $a_B = (\alpha m)^{-1}$ becomes equal to the  magnetic, or
Landau, radius $l_H = (eB)^{-1/2}$, the oscillator energy $eB/ 2m_e$ becomes
equal to Rydberg energy $Ry= m_e\alpha^2/2$. We use the system of units $\hbar
=c =1, \alpha= e^2=1/137$, dimensionless MF is defined as $H=B/B_a$. In this
system of units GeV$^2 =1.45\cdot 10^{19} G$.  The energy to change the
electron spin from antiparallel to parallel  to B is equal to 2H in units of
Rydberg. In terms of $H$ MF is classified \cite{11} as low $(H<10^{-3})$,
intermediate, also called strong $(10^{-3} < H <1)$, and intense ($1<H<\infty)$.
It seems natural to call MF $eB \sim \Lambda^2_{QCD}$ super-intense, and to say
that in this region ``QED meets QCD'' \cite{1}.

\section{Quarks in super-intense MF: a compendium of the results}

A number of papers published on this subject in recent years is of the order of
a hundred. Here we present in a very concise form the results of ITEP group
(M.A.Andreichikov, B.O.Kerbikov, V.D.Orlovsky and Yu.A.Simonov) \cite{12}.
Consider meson or baryon made of quarks embedded in strong MF. There are two
parameters defining the transition to the  regime when the mass and the
geometrical shape of the hadron undergo important changes. The first one is the
hadron size $r_h \simeq (0.6-0.8) $ fm. The strength of MF corresponding to it
is defined by $l_h \simeq r_h$ which yields $B_h \simeq 10^{18} G$. Another
related parameter is the string tension $\sigma \simeq 0.18$ GeV$^2$
responsible for the confinement. From the condition $eB/\sigma \simeq 1$ we
obtain $B_\sigma\simeq 10^{19} G$. It is therefore clear that the problem of
hadron properties in MF of the order of $(10^{18}-10^{19}) G$ has to be
formulated and solved at the quark level. The main questions is whether in
super-strong MF the hadron mass, e.g., that of the $\rho$- meson, falls down to
zero. For the quark system the question is whether MF induces the ``fall to the
center'' phenomenon. It  was shown by ITEP group that the answers to both
questions  are  negative.

The relativistic few-body problem is  hindered by well-known difficulties.
Maybe the most efficient method to solve the problem is the Field-Correlator
Method leading to the relativistic Hamiltonian \cite{12}.  To elucidate this
formalism   is beyond the scope of this presentation.  The method    includes
the following steps:

\begin{description}
    \item[a)] Fock-Feynman-Schwinger proper time representation of the Green's
function.
 \item[b)]Derivation of the confinement and OGE (color Coulomb) interactions using minimal surface
Wilson loop. \item[c)] Introduction of the quark dynamical masses (einbein
formalism).
 \item[d)]Inclusion of the  spin-dependent interactions $\vesig \veB$ and hyperfine.
 \item[e)] Derivation of the relativistic Hamiltonian $\hat H$ as the end-result of a)-d).
 \item[f)] Determination of the hadron mass and wave function
\end{description}

At step  b) one obtains the confinement interaction in the form $\sigma
|\ver_i-\ver_j|$ with $\sigma$ being the string tension. In order to obtain
analytical and physically transparent results we replaced the linear potential
according to \be V_{conf} = \sigma r \to \frac{\sigma}{2} \left(
\frac{r^2}{\gamma }+\gamma\right), \label{16}\ee
where $\gamma$ is a variational parameter. Minimizing (\ref{16}) with respect
to it, one retrieves the original form of $V_{conf}$. As was shown by the
numerical calculations, the accuracy of this procedure is $\la 5\%$. With the
account of MF and confinement, but without spin-dependent terms, the
hamiltonian has the form \be
\hat H_{q\bar q} = \frac{1}{\omega} \left( -
\frac{\partial^2}{\partial  \ver^2} + \frac{e^2}{4} (\veB\times \ver)^2\right)
+ \frac{\sigma}{2} \left( \frac{r^2}{\gamma} + \gamma\right).\label{16a}
\ee
This is a two-oscillator problem similar to (\ref{10})-(\ref{11}). We are
focusing on the ground state, hence the pseudomomentum  can be taken equal to
zero (see (\ref{11})), and it does not enter into (\ref{16a}). We note in
passing that in the relativistic Hamiltonian approach we evade a subtle problem
of the center-of-mass of the relativistic system. The mass eigenvalue
$M(\omega)$ and the dynamical mass $\omega$ are determined from a set of
equations.

\be \hat H \psi = M(\omega) \psi, ~~ \frac{ \partial M(\omega)}{\partial
\omega} =0.\label{17}\ee The wave function which is a solution of (\ref{17})
with the Hamiltonian (\ref{16a})  is \be \Psi(\ver) = \frac{1}{\sqrt{\pi^{3/2}
a^2_\bot a_z}}\exp \left( - \frac{r^2_\bot}{2a^2_\bot}-
\frac{r^2_z}{2a^2_z}\right),\label{18}\ee where at $e B\gg \sigma$ one has
$a_\bot \simeq \sqrt{\frac{2}{eB}}, ~~ a_z \simeq \frac{1}{\sqrt{\sigma}}$.
With MF increasing the $q\bar q$ system acquires the form of an elongated
ellipsoid, see Fig.~1.

\begin{figure}[h]
  \centering
  \includegraphics[width=0.7\textwidth]{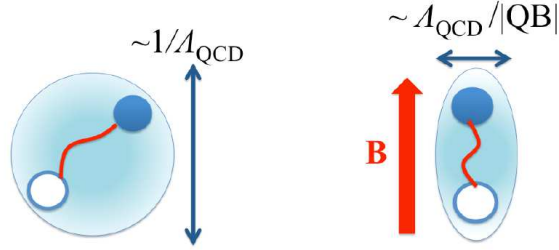}
    \caption{A sketch of MF influence on the meson wave function (from Ref.~\refcite{13}).}
\end{figure}


A similar behavior was observed before for the hydrogen atom in strong MF
\cite{14}. The difference is that  the longitudinal size of the $q\bar q$-meson
is bounded by $a_z \sim 1/\sqrt{\sigma}$ in contrast to the hydrogen atom which
in a strong MF takes the needlelike form with $a_z \sim (\ln H)^{-1}$.

The contribution of $V_{OGE}$ (color Coulomb) was calculated as the average
value of $ V_{OGE}$ over the wave function (\ref{18}) with quark and gluon loop
corrections  taken into account. Hyperfine (hf) spin-spin interaction was
treated in a similar way. Here a special care should be devoted to the $\delta$-function. Taken literally, it would lead to a divergent $\psi^2 (0) \sim eB$
factor (see the next section). Therefore the $\delta$-function was smeared over
the radius $\sim 0.2$ fm.

\begin{figure}[h]
  \centering
  \includegraphics[width=0.7\textwidth]{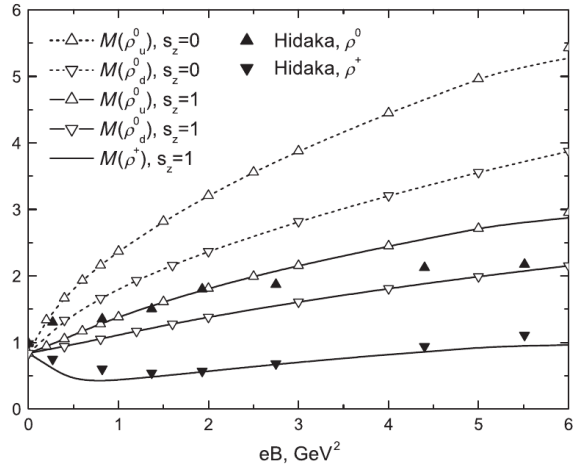}
    \caption{The results of the calculations of the $\rho$- meson mass as a
function of MF strength together with the lattice data \cite{15}.}
\end{figure}

In Fig. 2 the results for the $\rho$-meson mass as a  function of MF strength
ar presented together with the lattice data \cite{15}. We remind that MF
violates  both spin and isospin symmetries. In order to minimize the Zeeman
energy the lowest state of $u\bar u$ (or $d\bar d$) in strong MF becomes spin
polarized $|u\uparrow \bar u\downarrow \ran$. In  our somewhat oversimplified
picture this state is a mixture of  $\rho^0$ and $\pi^0$. The conclusion is
that the mass of the quark-antiquark state does not reach zero no matter now
strong MF is. The same result is true for the neutron made of three quarks.

Here we covered only few results of ITEP group on quarks in MF --- see Ref.~\refcite{12}.

\section{The new results on Zeeman levels in hydrogen}

The spectrum of hydrogen atom (HA) in MF is  a classical problem described
textbooks \cite{4}. The present wave of interest to superstrong MF inspired the
reexamination of this problem \cite{16,17,18,19}. Surprisingly enough, the new
important results were obtained. It was shown that in superstrong  MF radiative
corrections  screen the Coulomb potential thus leading to the  freezing of the
ground state energy at the value $E_0 = -1.7$ keV \cite{18, 19}.

Here we discuss the new correction to hyperfine (hf)  splitting in HA
\cite{14}. In  HA the dramatic changes of the ground state binding energy and
the wave function geometry occurs starting from $H\simeq 1$. In this region
magnetic confinement in the  plane perpendicular to MF dominates the Coulomb
binding to the proton. With MF strength growing, the binding energy rises
\cite{16,17,18,19}. The wave function squeezes and takes the needlelike form.
The probability to find the electron  near the proton increases. This means
that the value of the wave function at the origin $|\psi (0)|^2$ depends on MF
and in the limit $H\gg 1$ one has $|\psi(0)|^2\sim H \ln H $ \cite{14}. This
phenomenon may be called ``Magnetic Focusing of Hyperfine Interaction in
Hydrogen''. In addition, the deviation of the HA ground state wave function
from the spherical symmetry results in the appearance  of the tensor forces.
These two MF induced effects result in corrections to the standard picture (see
Fig.~3) of the Zeeman splitting. The  energies of the splitted levels are found
by the diagonalization of the following Hamiltonian \cite{14}

\begin{figure}[h]
  \centering
  \includegraphics[width=0.7\textwidth]{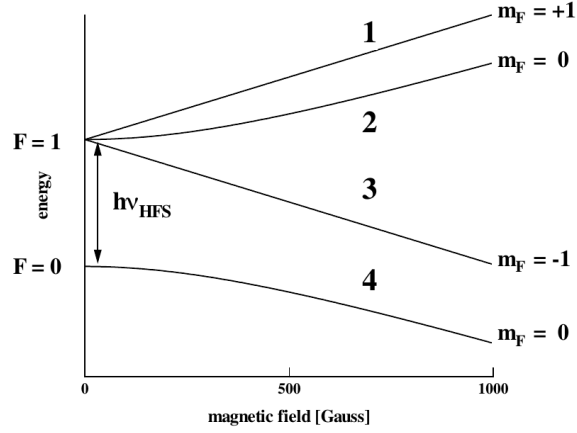}
    \caption{Hydrogen hf structure. Transition $|2\ran \to |4\ran$ at $B=0$
corresponds to the $\lambda=21$ cm (1420 MHz) line.}
\end{figure}


\be \hat H_{hf} =\frac{8\pi}{3} g\mu_B\mu_N |\psi_B (0)|^2 \vesig_e \vesig_p +
\frac{8\pi}{3} \delta \psi_B(0) \sigma_{ez} \sigma_{pz} + \mu_B
(\vesig_e\veB) - g\mu_N (\vesig_p\veB).\label{19}\ee

Here $g=2.79, g\mu_n$ is the proton magnetic moment, $\mu_B$ is the Bohr
magneton, index $B$ affixed to $\psi_B$ and $\delta \psi_B$ indicates the
dependence on MF. At $B=0$ one retrieves the standard expression with
$|\psi_B(0)|^2 = m^3 \alpha^3/ \pi$ and $\delta \psi_B (0) =0$. We do not
discuss the corrections to the $\Delta E_{hf} = 1420 MHz$ line due to
relativistic effects, QED, and nuclear structure. This  subject is thoroughly
elucidated  in the  literature \cite{20,21}. For the frequency of the $\Delta
F=1, \Delta m_F=0 ~~|2> \to |4>$ transition diagonalization of the Hamiltonian
(19) yields \be \nu=E_2 - E_4 = \Delta E_{hf} \sqrt{\gamma^2 + \left(
\frac{2\mu_BB}{\Delta E_{hf}} \right)^2\left( 1+
g\frac{m_e}{m_p}\right)^2}.\label{20}\ee

 Here $\gamma^2$ is a new MF dependent
parameter with the following asymptotic behavior \cite{14}. \be \gamma\to
1+\left( 1- \frac{a^2_\bot}{a^2_z}\right), ~~ H\ll \alpha^2 \frac{m_e}{m_p}
\simeq 10^{-7},\label{21}\ee \be \gamma \to H\ln H, ~~ H\gg 1, \label{21}\ee
\be \gamma\gg \frac{2\mu_BB}{\Delta E_{nf}},~~ \ln H\gg10^7.\label{23}\ee

In the standard picture without magnetic focusing $\gamma \equiv 1$.

The question is whether magnetic focusing in HA can be experimentally detected
in the laboratory conditions. A very preliminary positive answer relies on
extremely accurate experiments in search of Zeeman frequency variation using
the hydrogen maser \cite{22}. It typically operates with constant MF of the
order of $\sim 1$ mG. In this regime the frequencies of the $F=1, \Delta
m_F=\pm 1$ Zeeman transitions were measured with a precision of $\sim 1$ mHz
\cite{22}. This subject deserves  a detailed discussion to  be presented in
another publication.

This presentation is based on the work of the ITEP team: M.~A.~Andreichikov,
B.~K., V.~D.~Orlovsky and Yu.~A.~Simonov.

The author gratefully acknowledges the encouraging discussions with
M.~I.~Vysotsky, S.~I.~Godunov, V.~S.~Popov, B.~M.~Karnakov, A.~E.~Shabad and
A.~Yu.~Voronin.


\begin{thebibliography}{99}


\bibitem{1}

D. E. Kharzeev, K. Landsteiner, A. Schmitt, and H.-U. Yee, Lect. Notes Phys.
\textbf{871}, 1 (2013).
\bibitem{2}
 D. E. Kharzeev, L. D. McLerran and H. J. Warringa, Nucl. Phys.A {\bf 803}, 227
 (2008); V. Skokov, A. Illarionov and V. Toneev, Int. J. Mod. Phys.A {\bf 24},
 5925 (2009).
\bibitem{3} A. Y. Potekhin, Phys. Usp. \textbf{53}, 1235 (2010);  A.~K. Harding and Dong Lai, Rept. Prog. Phys. \textbf{69}, 2631 (2006).

\bibitem{4} L. D. Landau and E. M. Lifshitz, Quantum mechanics. Course of
Theoretical Physics, vol. 3, Pergamon Press, Oxford (1978).


\bibitem{5} W. E. Lamb, Phys. Rev. {\bf 85}, 259 (1952); L. P. Gor'kov and
I. E. Dzyaloshinskii, Soviet Physics JETP, {\bf 26}, 449 (1968); J. E. Avron,
I. W. Herbst, and B. Simon, Ann. Phys. (NY), {\bf 114}, 431 (1978); H. Grotsch
and R. A. Hegstrom, Phys. Rev. A {\bf 4}, 59 (1971).

\bibitem{6} H. Herold, H. Ruder, and G. Wunner, J. of Phys. B \textbf{14}, 751 (1981).
\bibitem{7} Dong Lai, Rev. Mod. Phys. {\bf 73}, 629 (2001).

\bibitem{8} J. Alford and M. Strickland, Phys. Rev. D {\bf 88}, 105017 (2013).

\bibitem{9} Yu. A. Simonov, Phys. Lett. B \textbf{719}, 464 (2012).
\bibitem{10} A. Yu. Voronin and P. Froelich, Phys. Rev. A {\bf 77},  022505
(2008).

\bibitem{11} M. D. Jones, G. Ortiz, and D. M. Ceperley, Phys. Rev. A {\bf 54}, 219
(1996).


\bibitem{12} M.~A. Andreichikov, B.~O. Kerbikov, V.~D. Orlovsky, and Yu.~A. Simonov, Phys. Rev. D \textbf{87}, 094029
(2013);
 M.~A. Andreichikov, V.~D. Orlovsky, and Yu.~A. Simonov, Phys. Rev. Lett. \textbf{110}, 162002 (2013);
 V.~D. Orlovsky and Yu.~A. Simonov, JHEP 1309, 136 (2013), arXiv:1306.2232 [hep-ph];
 Yu.~A. Simonov, Phys. Rev. D \textbf{88}, 025028 (2013), arXiv:1303.4952
[hep-ph]; M.~A. Andreichikov, B.~O. Kerbikov, Yu.~A. Simonov, arXiv:1304.2516
[hep-ph];Yu.~A. Simonov, arXiv:1308.5553 [hep-ph]; Yu.~A. Simonov, Phys. Rev. D
\textbf{88}, 053004 (2013).

\bibitem{13} Toru Kojo and Nan Su, The quark mass gap in  a magnetic field,
arXiv:1211.7318 [hep-ph].

\bibitem{14} M.~A. Andreichikov, B.~O. Kerbikov, and Yu.~A. Simonov, Magnetic field focussing of hyperfine
interaction in hydrogen, arXiv:1304.2516 [hep-ph].

\bibitem{15} Y. Hidaka and A. Yamamoto, Phys. Rev.  D {\bf 87}, 094502 (2013).

\bibitem{16} B. M. Karnakov, V. S. Popov, J. Exp. Theor. Phys. {\bf 114}, 1 (2012);
Physics uspekhi, accepted for publication.

\bibitem{17} A. E. Shabad and V. V. Usov, Phys. Rev. Lett. {\bf 98}, 180403 (2007);
Phys. Rev. D {\bf 77},  025001 (2008).

\bibitem{18} B. Machet and M. I. Vysotsky, Phys. Rev. D {\bf 83}, 025022 (2011).

\bibitem{19} S. I. Godunov, B. Machet, M. I. Vysotsky, Phys. Rev. D {\bf 85}, 044058
(2012).

\bibitem{20}  S. G. Karshenboim, Phys. Rept., {\bf 422}, 1 (2005).

\bibitem{21} M.~I. Eides, H. Grotch, and V. A. Shelyuto, Phys. Rept., {\bf 342}, 63
(2001).

\bibitem{22}
 D.~F. Phillips \textit{et al.}, Phys. Rev. D {\bf 63}, 111101 (2001);  M.~A. Humphrey \textit{et
 al.}, Phys. Rev. A {\bf 68}, 063807 (2003).
\end{thebibliography}
\end{document}